\documentclass[twocolumn, prb, superscriptaddress,
  showpacs]{revtex4-1} 
\usepackage{amsmath, array} 
\usepackage{graphicx}
\usepackage{grffile}
\usepackage{hyperref} 
\usepackage{color}
\usepackage{xcolor}

\begin{document}
\title{Quantum transport in an environment parametrized  by dispersive bosons}
\author{Monodeep Chakraborty} \email{bandemataram@gmail.com} \affiliation{Centre for Quantum Science and Technology, Chennai Institute of Technology, Chennai-600069, India}
\author{Holger Fehske}\email{fehske@physik.uni-greifswald.de}
	\affiliation{Institute of Physics, University of Greifswald, Felix-Hausdorff-Straße 6, 17489 Greifswald, Germany}
	\affiliation{Erlangen National High Performance Computing Center, 91058 Erlangen, Germany}
\begin{abstract}
We  generalize the two-channel (Edwards) fermion-boson model describing quantum transport in a background medium to the more realistic case of dispersive bosons. Using the variational  exact diagonalization technique, we numerically solve the extended model  in a one-dimensional setting, for both downward and upward curved boson dispersion, and show that going away from the previous Einstein-boson assumption has profound  consequences for the particle transport. Specifically, we analyze the ground-state and spectral properties and demonstrate the renormalization of the particle's coherent band, effective mass, 
photoemission spectra and fermion-boson correlation functions 
due to the interplay of fluctuations and correlations in a dispersive environment.    
\end{abstract}
\date{\today}
\maketitle

\section{Introduction}
As a particle moves through a system, it is constantly interacting with some background medium. In this way, the transport is strongly influenced by the correlations and fluctuations present in the environment. 
The situation becomes more complicated in so-called strongly correlated systems, where the properties of the background are determined by the motion of the particle itself. Such systems have been of major interest in condensed matter physics over the last few decades. Perhaps the most prominent examples are quasi-1D (one-dimensional) halogen-bridged transition metal complexes\cite{BS93},  quasi-2D (two-dimensional) high-$T_c$ superconducting cuprates\cite{BM86}, and  three-dimensional (3D) colossal magnetoresistive manganites\cite{JS50}. In all these cases, the striking transport properties appear when doping the insulating parent compounds exhibiting charge, spin, or orbital order\cite{KLR89,MH91a,WOH09,Be09}. Interestingly, coherent transport can evolve in such materials regardless of  strong background correlations, albeit on a reduced energy scale.  

The complexity of the electron-electron interaction effects usually prevents the analytical or even numerical solution of the quantum many-particle models commonly used for the theoretical description of these systems. A way out may be to consider instead simplified transport models where the particle motion takes place in an effective background medium with spin, orbital or lattice degrees of freedom that are parametrized by bosons. The Edwards fermion-boson constitutes a paradigmatic model in this respect\cite{Ed06}. It describes a very general situation: As a particle moves along a 1D transport path, it affects the adjusted background by creating a bosonic excitation of a certain energy at the sites it leaves (or annihilating an existing energy at the site it enters), but of course any distortion of the background is allowed to relax by quantum fluctuations. Thus a large (small) boson energy and a small (large) boson relaxation rate hinder (help) transport. In one dimension, the Edwards model has been solved in the one-particle sector by variational exact diagonalization, where different (quasi-free, diffusive and boson-assisted) transport regimes have been identified\cite{AEF07}. For the half-filled band case, a Luttinger liquid charge-density-wave (metal-insulator) quantum phase transition was proved to exist by density-matrix renormalization group calculations\cite{EHF09,EF09b}.  The formation of Edwards polarons has been discussed in 2D and 3D\cite{BF10,CMTMF16}.

In deriving the Edwards fermion-boson model, it was assumed that the background parametrizing bosons are dispersionless\cite{Ed06}, as are the optical lattice phonons in the Holstein small polaron model\cite{Ho59a}.  However, this simplification must be questioned\cite{RL84}. Recent studies  of an appropriately modified single-particle Holstein model have provided evidence of this, since the polaron transport changes significantly when the phonons become non-local\cite{MB13,BT21}. For the half-filled Holstein model, the importance of a finite phonon bandwidth and curvature on the competition between pairing and charge order has been worked out\cite{CBCBS18}.  To what extent dispersive bosons will  influence the transport behavior of the Edwards model is a completely open question and will be the main topic of this work. 

We proceed as follows. In Sec.~II we introduce the generalized Edwards fermion-boson model and briefly outline  the scheme for its numerical solution. In Sec. III we present and analyze the ground-state and spectral properties of the model and discuss in particular the effects of the dispersive boson on quantum transport. Our conclusions are given in Sec. IV.

\section{Model and Method}
We consider a single-particle coupled to bosons on an infinite 1D lattice:
\begin{align}\label{Ham}
H =&  - t_f \sum_{\langle i, j \rangle}  f_j^{\dagger} f^{}_i 
-t_{fb} \sum_{\langle i, j \rangle}  f_j^{\dagger} f^{}_i (b_i^{\dagger}
+b^{}_j) \\&+ t_b \sum_{\langle i, j \rangle}  b_j^{\dagger} b^{}_i + \omega_0 \sum_i b_i^{\dagger} b^{}_i \,.
\end{align}
The Hamiltonian Eq.~\eqref{Ham} models the two transport channels realized in many condensed matter systems:
coherent particle transfer on an (often very reduced) energy scale $\propto t_f$ and boson-affected (or perhaps even boson-controlled) 
hopping with amplitude  $t_{fb}$. Each time a fermion $f_i^{(\dagger)}$ hops it creates (absorbs) a boson $b_i^{(\dagger)}$
at the site it leaves (enters).  The bosons are assumed to be dispersive, forming a band with center $\omega_0$, i.e., 
\begin{equation}
\omega(q)=\omega_0 +2 t_b \cos(q)\;.
\label{PD}
\end{equation}
In this way, $H$ mimics the correlations or fluctuations of a spinful fermionic many-body system by
spinless particles with boson-influenced hopping. Note that shifting the boson operators  $b_i\to b_i - \tfrac{\lambda}{\omega_0}$, where $\lambda=\tfrac{\omega_0}{2}\tfrac{t_f}{t_{fb}}$ (cf. Refs.[~\onlinecite{AEF07,EF09b}]), will replace the first term of Eq.~\eqref{Ham} 
by a boson relaxation term that depends on $\lambda$ and $t_b$.

The pure Edwards model describes an extremely complex transport behavior depending on the relative strength of the model parameters; compare the schematic phase diagram (Fig.~1) in Ref. [~\onlinecite{AEF07}].  For large $t_f/t_{fb}$ transport takes place through unrestricted hopping, where for large $\omega_0/t_{fb}$ the number of bosons is small and the particle 
propagates almost coherently. In the case of small $\omega_0/t_{fb}$, on the other hand, the number of bosons is noticeably larger and they appear as random scatterers, which leads to 
loss of coherence. In contrast, for small $t_f/t_{fb}$, the main transport process is boson-assisted hopping, i.e., the particle motion relies on the existence of bosons and a closely linked 
fermion-boson dynamics. In this regime  strong correlations develop in particular for large ratios $\omega_0/t_{fb}$. How a finite dispersion of bosons influences this scenario is still unclear
and must be examined in the entire parameter range.

In order to determine the ground-state (static) and spectral (dynamic) properties of the modified Edwards model Eq.~\eqref{Ham}   in the one-particle sector, we use a variant of the variational exact diagonalization (VED) scheme, which has been described in detail in Refs. [~\onlinecite{BTB99,KTB02,FT07,AEF07,CMTMF16}], and [~\onlinecite{CCF23}]. In a nutshell, the variational Hilbert 
space is generated starting from the initial single-electron Bloch state with a momentum $k$ in the first Brillouin zone of an infinite chain, $|k \rangle \propto \sum_j \exp(i k R_j) f^\dagger_j|{\rm vac}\rangle$, where  $|{\rm vac}\rangle$ is the vacuum state with no bosons. Further basis states are generated applying the first two  off-diagonal (electron-hopping) terms  in Eq.~\eqref{Ham} $N_{gen}$ times. Furthermore all the translations of these states on an infinite lattice are  taken into account, i.e., we can calculate quantities at fixed momenta $k$.  
As a result of this procedure, the maximum boson number at the electron position is $N_{gen}$. At the same time, the maximum distance of a boson from the electron site is $N_{gen}-1$. The finite $N_{gen}-1$ leads to a discrete boson dispersion $\omega(q)$\cite{BT21}. When calculating static quantities, we make sure an accuracy of at least 10 digits in the whole parameter regime. For this  $N_{gen}=16$ is usually sufficient. Calculating spectral quantities we have used an artificial broadening $\eta=0.01$\cite{WWAF06}.  In view of the above discussion of the relevant model parameter ratios, measuring all energies and frequencies in units of $t_{fb}$ is a very natural choice\cite{EHF09,EF09b}.

\section{Results and Discussion}

\subsection{Static properties}
We begin by analyzing the ground-state properties  of the 1D Edwards model with dispersive bosons.  
With regard to the physically most relevant applications, such as, e.g., lattice and spin polaron formation in strongly correlated electron systems,
ratios of $t_f/t_{fb}\leq 1$ are certainly most interesting.

\begin{figure}[t]
\includegraphics[width=0.95\linewidth]{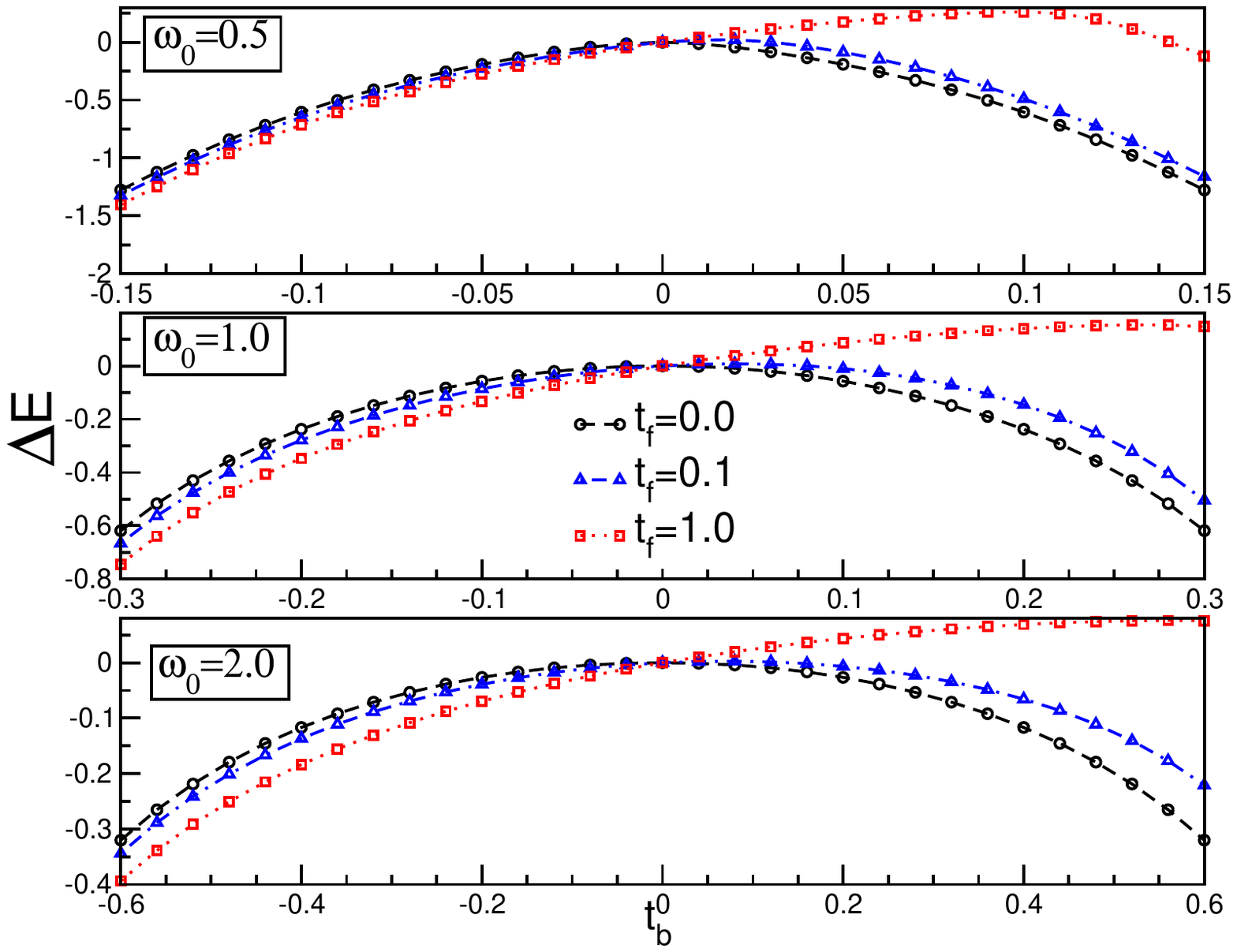} 
\caption{Ground-state energy of the 1D Edwards model with dispersive bosons centered around $\omega_0$.
Shown is the difference $\Delta E=E(0)-E_{t_b=0}(0)$ as a function of the boson transfer amplitude $t_b$ for $\omega_0=0.5$, 1.0, and 2 at various $t_f$.}
\label{fig1}
\end{figure}

Figure~\ref{fig1} shows the energy $E(0)$ in the  ground state  $|0\rangle$ with total momentum $k=0$, as a function of the boson transfer amplitude $t_b$ for characteristic values of $t_f$ and $\omega_0$ discussed in previous studies of the pure Edwards fermion-boson model. Considering the effect of the finite boson dispersion, we should keep in mind that $t_b>0$  ($t_b<0$)  increases (decreases) $\omega(0)$ compared to $\omega_0$ in Eq.~\eqref{PD}, and  bends $\omega (q)$  downward (upward) moving away from $q=0$. This limits the meaningful parameter values to $|t_b|< \omega_0/2$. 
Of course, the energy difference $\Delta E=E(0)-E_{t_b=0}(0)$ is always negative, and symmetric with regard to $t_b \leftrightarrow -t_b$, if the coherent particle hopping channel  disappears ($t_f=0$). At finite $t_f$,  the concave boson dispersion ($t_b<0$) is energetically more favorable in the ground state than the convex one  ($t_b>0$), mainly due to the lower value of $\omega (0)$. 
Especially for quasi-free particle transport, i.e., large values of $t_f$, $\Delta E$  is even positive for $t_b>0$. The asymmetry observed here and in the following figures with respect to $t_b\leftrightarrow -t_b$ at finite $t_f$ mainly results from the mixing between predominantly fermionic  states stemming from an always upward-bended band and predominantly bosonic states coming from an upward- ($t_b<0$) or downward-bended ($t_b>0$) bended dispersion, which, in the latter case, is less (more) pronounced  at small (large) momenta.

\begin{figure}[h!]
\includegraphics[width=0.95\linewidth]{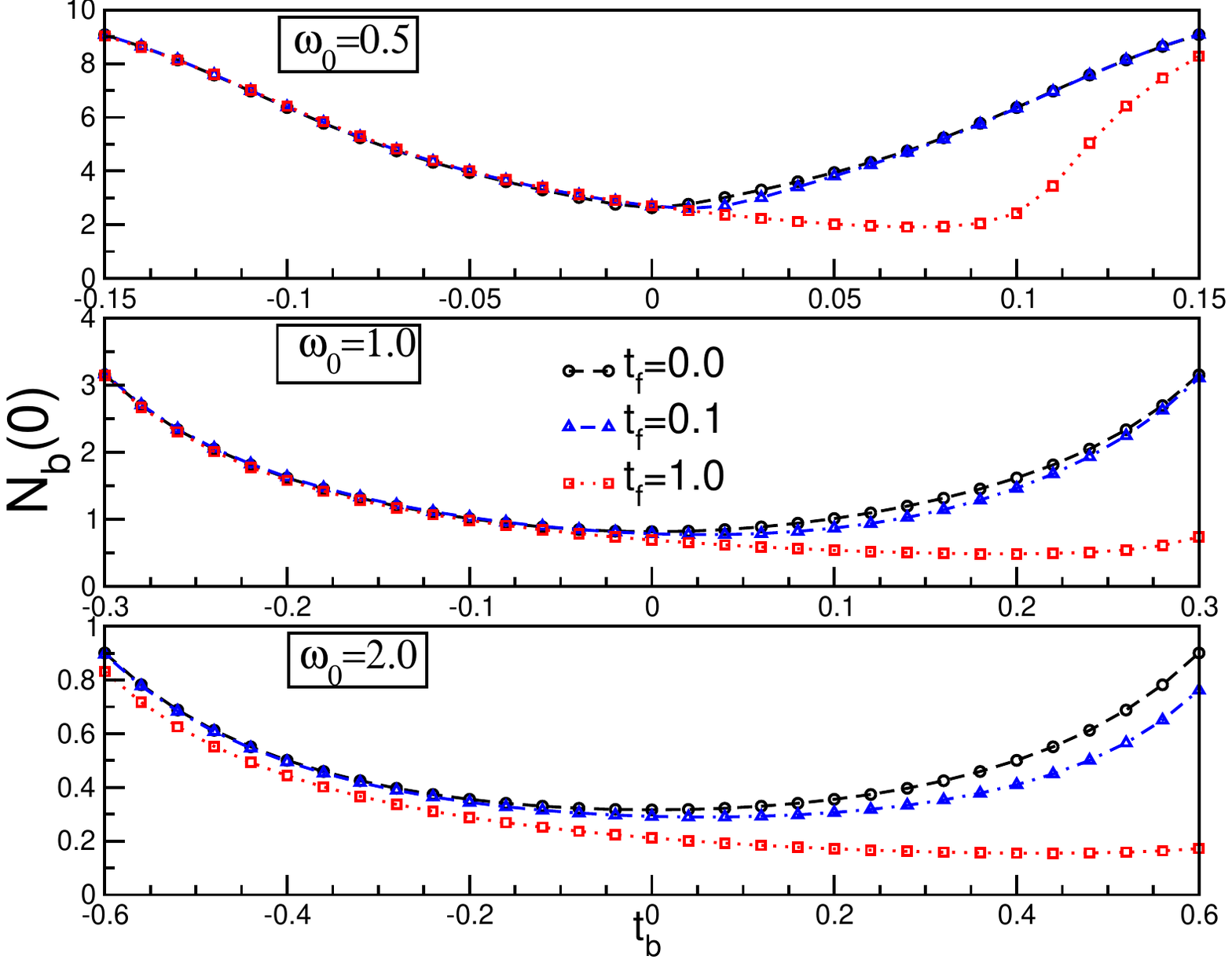} 
\caption{Mean boson number in the ground state of the 1D Edwards model with dispersive bosons centered around $\omega_0$.}
\label{fig2}
\end{figure}
\begin{figure}[h!]
\includegraphics[width=0.95\linewidth]{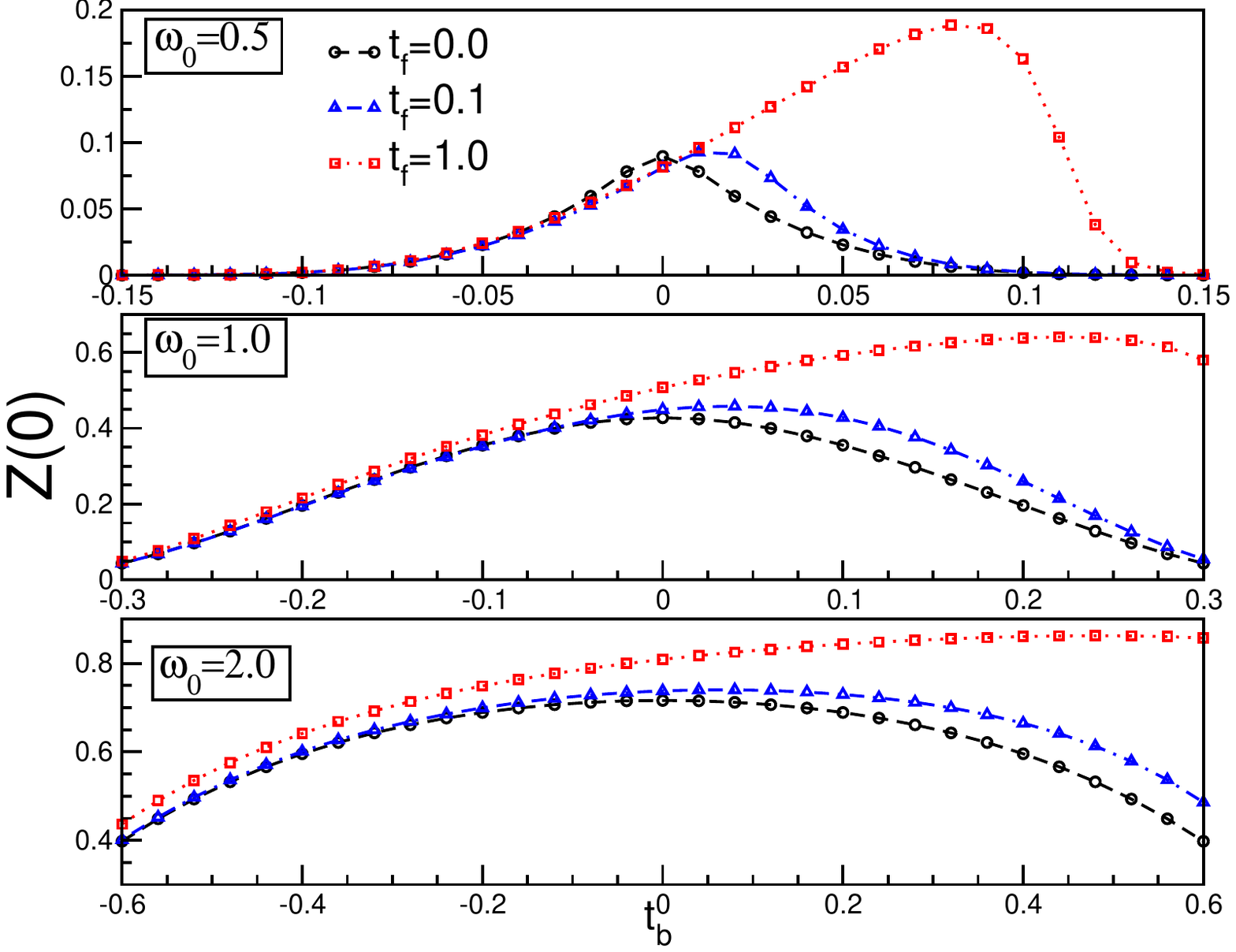} 
\caption{Quasiparticle weight in the ground state of the 1D Edwards model with dispersive bosons centered around $\omega_0$.}
\label{fig3}
\end{figure}
\begin{figure}[t]
\includegraphics[width=0.95\linewidth]{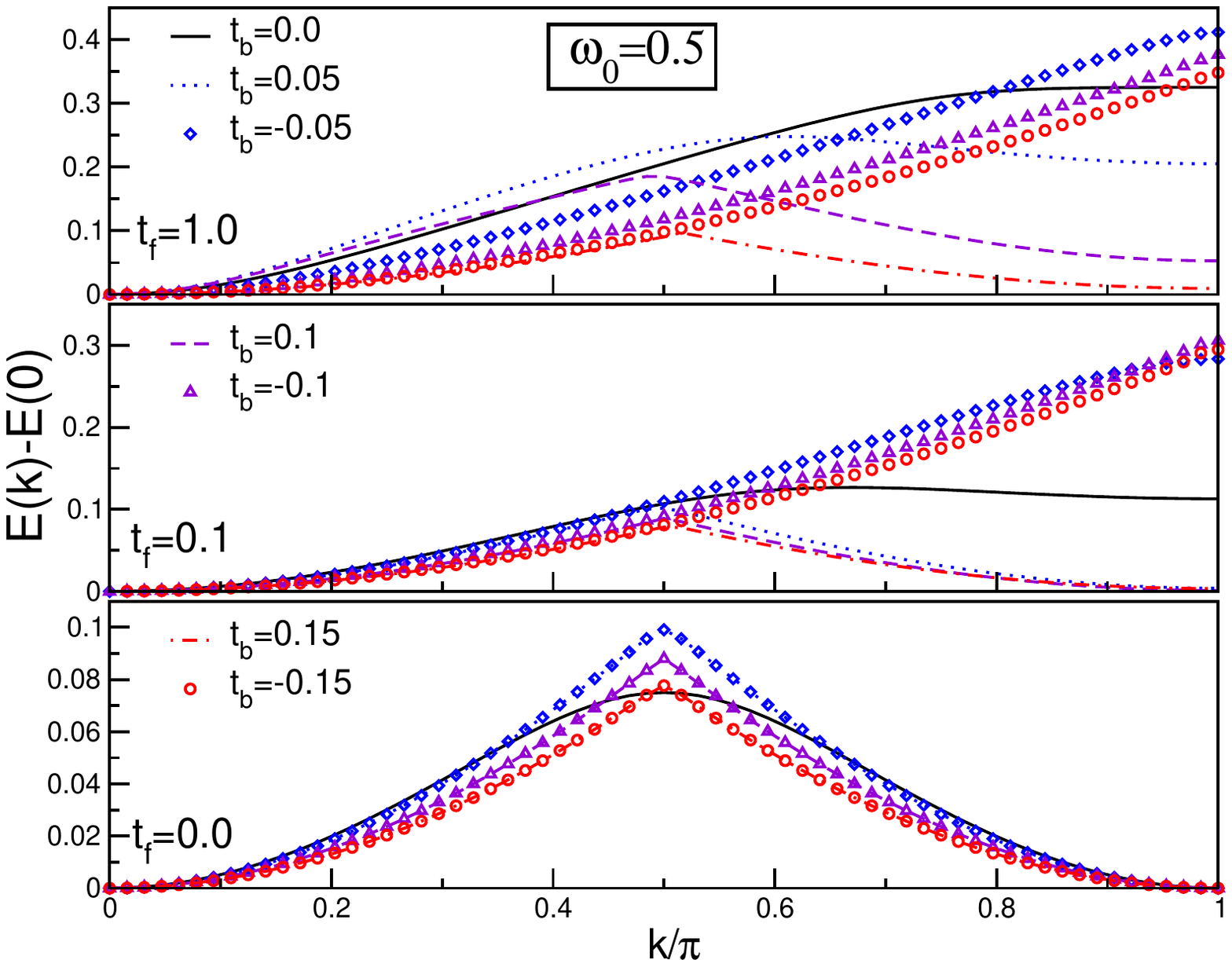} 
\caption{Lowest-energy bands of the 1D Edwards model with dispersive bosons centered around $\omega_0=0.5$. The black solid lines give the corresponding data of the regular Edwards model ($t_b=0$) for comparison. Take note of the different ordinates scaling.}
\label{fig4}
\end{figure}
\begin{figure}[t]
\includegraphics[width=0.95\linewidth]{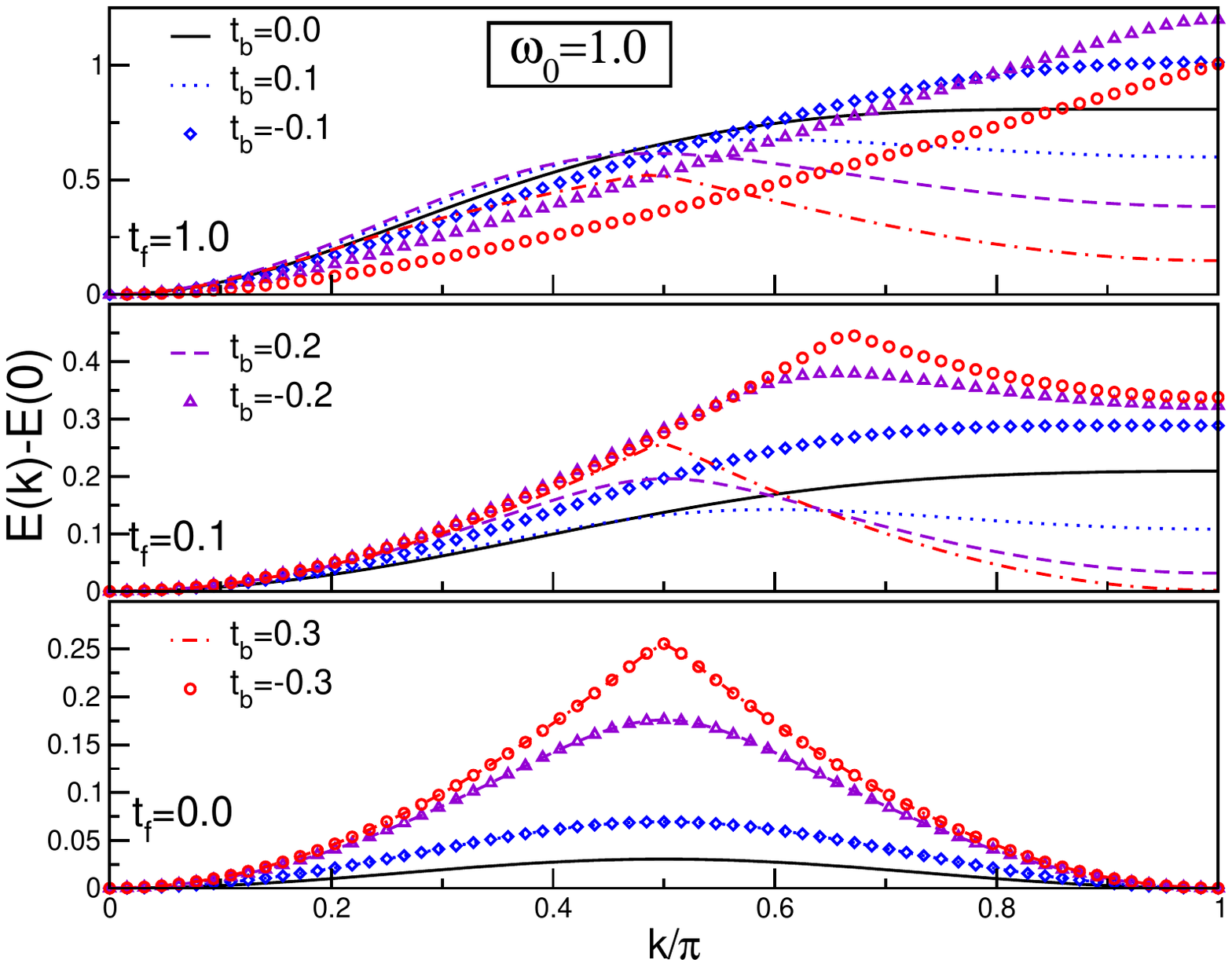} 
\caption{Lowest-energy bands of the 1D Edwards model with dispersive bosons centered around $\omega_0=1.0$. }
\label{fig5}
\end{figure}
\begin{figure}[t]
\includegraphics[width=0.95\linewidth]{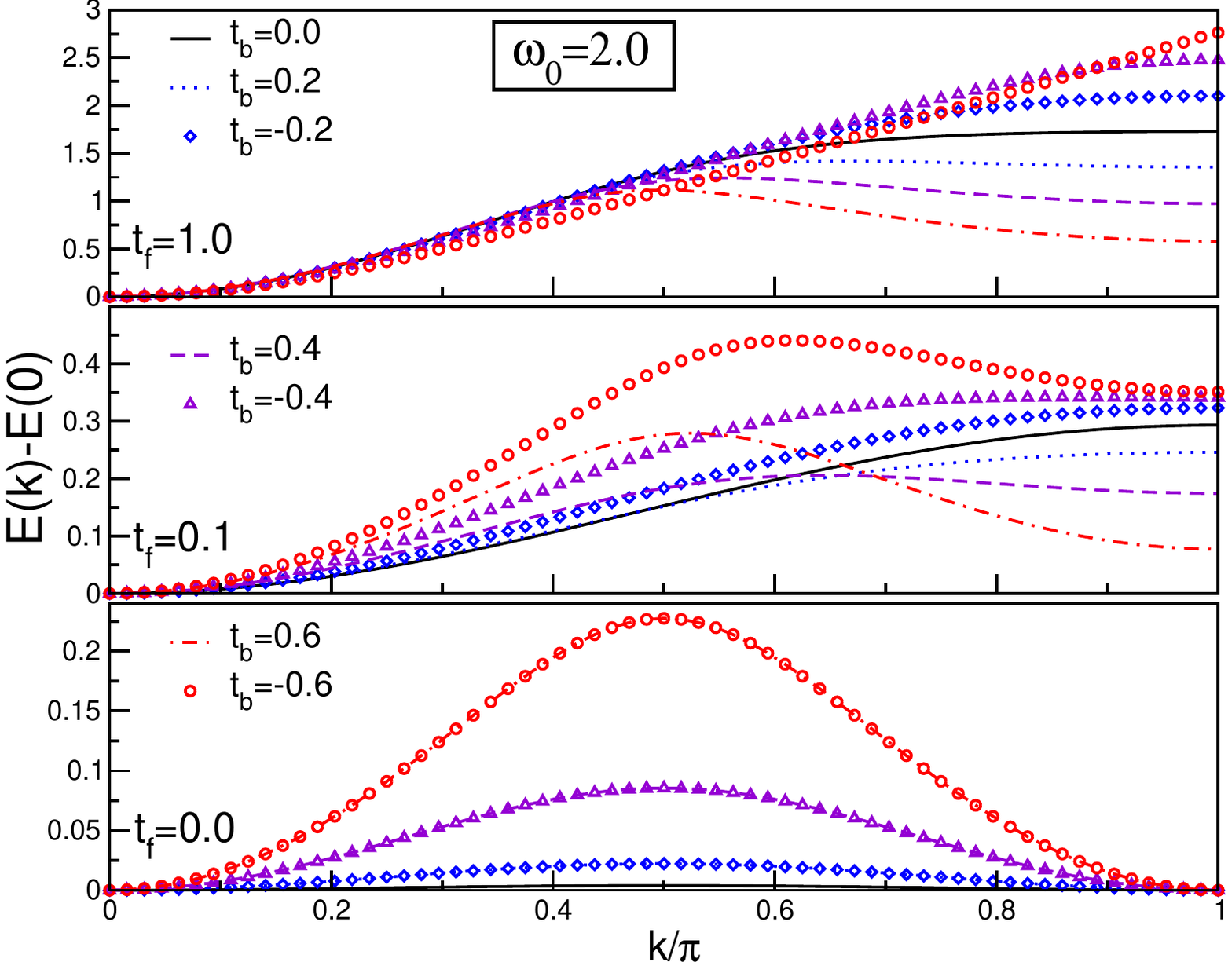} 
\caption{Lowest-energy bands of the 1D Edwards model with dispersive bosons centered around $\omega_0=2.0$. }
\label{fig6}
\end{figure}

To further characterize the nature of the ground state, we give in Figs.~\ref{fig2} and~\ref{fig3} the mean phonon number  $N_b(0)=\sum_i \langle 0|b_i^\dagger b_i^{}|0\rangle$  and the (electronic) quasiparticle weight $Z(k=0)=\vert\langle 0 \vert f_{k=0}^{\dag} \vert \mathrm{vac}\rangle\vert^{2}$, respectively, for the same model parameters as in Fig.~\ref{fig1}. Both figures provide a very consistent picture of the physics of the modified Edwards model. In the fluctuation-dominated (correlation-dominated) region of low (high) frequencies $\omega_0$ and large (small) $t_f$, where bosonic excitations of the background cost little (much) energy, the average number of bosons in the ground state is rather high (low). The quasiparticle weight is relatively low (high) accordingly. Dispersive bosons generally increase the possibility of triggering excitations in the background medium that affect the particle transport. These bosons act as random (incoherent) scatters. In line with this, $N_b(0)$ grows and $Z(0)$ becomes smaller. An exception is the case of almost-free particles, i.e., large $t_f$  and positive $t_b$, where the number of bosons (the quasiparticle weight) decreases (increases) with increasing $t_b$, since the bosons are energetically more expensive. Here the number of bosons only increases significantly when the boson dispersion is bent far enough downwards due to the positive $t_b$. Obviously this effect is more pronounced the smaller the $\omega_0$ is. In the limit of small $t_f$ and large $\omega_0$, when transport is boson-assisted, a finite $t_b$ acts in the opposite direction, i.e., $N_b(0)$ is rather small and consequently $Z(0)$ is getting bigger (note the different scales of the ordinates).

In Figs.~\ref{fig4}--\ref{fig6} we present the energy band $E(k)$ that we get from the lowest eigenvalue of $H$ in each $k$ sector 
at  different values of $\omega_0$, $t_f$, and $t_b$. Let us emphasize once more that $k$ refers to the total momentum of the coupled fermion-boson system.

If we first look at the results of the pure Edwards model ($t_b\equiv 0$, black solid lines), we see that 
the band dispersion is only weakly renormalized for  $t_f$'s in the order of $t_{fb}$ or larger,
with the exception of the band flattening at larger momenta when the (dispersionless) boson crosses 
the electronic cosine band, provided $\omega_0$ is not too large. In contrast, the quasiparticle bandwidth becomes 
tiny for small $t_f$, especially if $\omega_0$ is large, i.e., when the background medium is stiff. 
Here, the particle transport is fully boson assisted. This can be best understood in the limit $t_f=0$, 
where the lowest-order vacuum-restoring process  comprises six steps and thereby propagates the 
particle by two sites\cite{AEF07}. In this case, $E(k)$ has the period $\pi$, 
see the bottom panels of Figs.~\ref{fig4}--\ref{fig6}.

The inclusion of the boson dispersion significantly affects the band structure of the lowest-energy band. In the quasi-free transport regime of large $t_f$ and for positive values of $t_b$, the dispersion initially largely follows that of the pure Edwards model but then bends down for large momenta where the (bare) phonon dispersion cuts into the electronic one. This effect is enhanced when $t_b$ grows and somewhat more pronounced for small $\omega_0$. In contrast, for negative $t_b$, the dispersion grows more or less continuously.  The difference
 between the convex and concave boson-dispersion cases is less distinctive and finally vanishes when strong correlations develop in the system, i.e., for small values of $t_f$ and large 
 values of $\omega_0$. At the same time it turns out that the particle band becomes more dispersive in the boson-controlled transport regime $t_f \simeq 0,\, \omega_0  >1$, again compared to the case $t_b=0$ (pure Edward model), see lower panels. This means that dispersive bosons increase the mobility of the particles in this regime, regardless of whether there is a positive or negative curvature.   
 
\begin{figure}[t]
\includegraphics[width=0.95\linewidth]{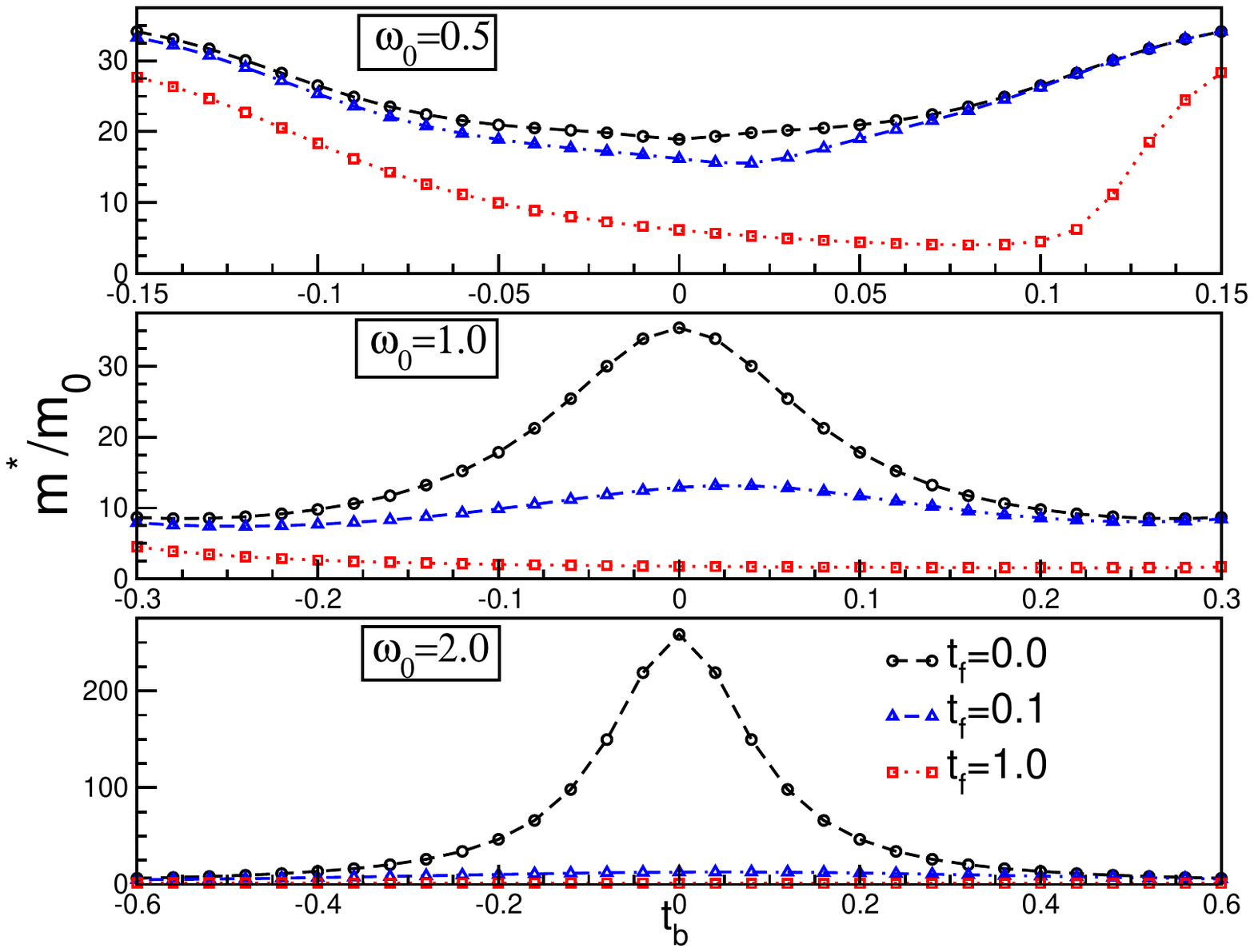} 
\caption{Mass enhancement in the 1D Edwards model with dispersive bosons as a function of $t_b$. Take note of the different ordinates scaling.}
\label{fig7}
\end{figure}

Figure~\ref{fig7} shows the effective particle mass in the modified Edwards model, 
\begin{equation}
m^* = \left[\left. \frac{\partial^2 E(k)}{\partial k^2}\right|_{k=0}\right]^{-1}\,,
\end{equation} 
more precisely  the mass enhancement due to the coupling to the background bosons compared to the bare band mass $m_0=1/2t_{fb}$. 
Note that in the Edwards model, the Drude weight is related to the effective mass by Kohn's formula\cite{Ko64},
\begin{equation}
D = 1/(2m^*)\,,
\end{equation} 
where $D$ serves as a measure of coherent transport\cite{AEF07,AEF10}.
Again, the influence of the boson dispersion is very different in the fluctuation and correlation dominated regimes. At low boson frequencies, the larger number of bosons---appearing in the fluctuating background for larger values of $t_b$---causes a moderate increase of the quasiparticle mass. In contrast, for high boson frequencies only a few bosons exist in the background, which is stiff, i.e., strongly correlated, and consequently there is a huge mass enhancement when $t_f$ is small. In this case, the additional bosons  showing up for $|t_b|>0$ lead to a significant reduction of the particle's  effective mass  because they promote hopping processes. Of course, a distinct quasi-free transport channel, as it appears also for large values of $\omega_0$ provided that $t_f$ is large as well, shows a rather weak dependence on $t_b$, see bottom panel.   

\begin{figure}[t]
\includegraphics[width=0.95\linewidth]{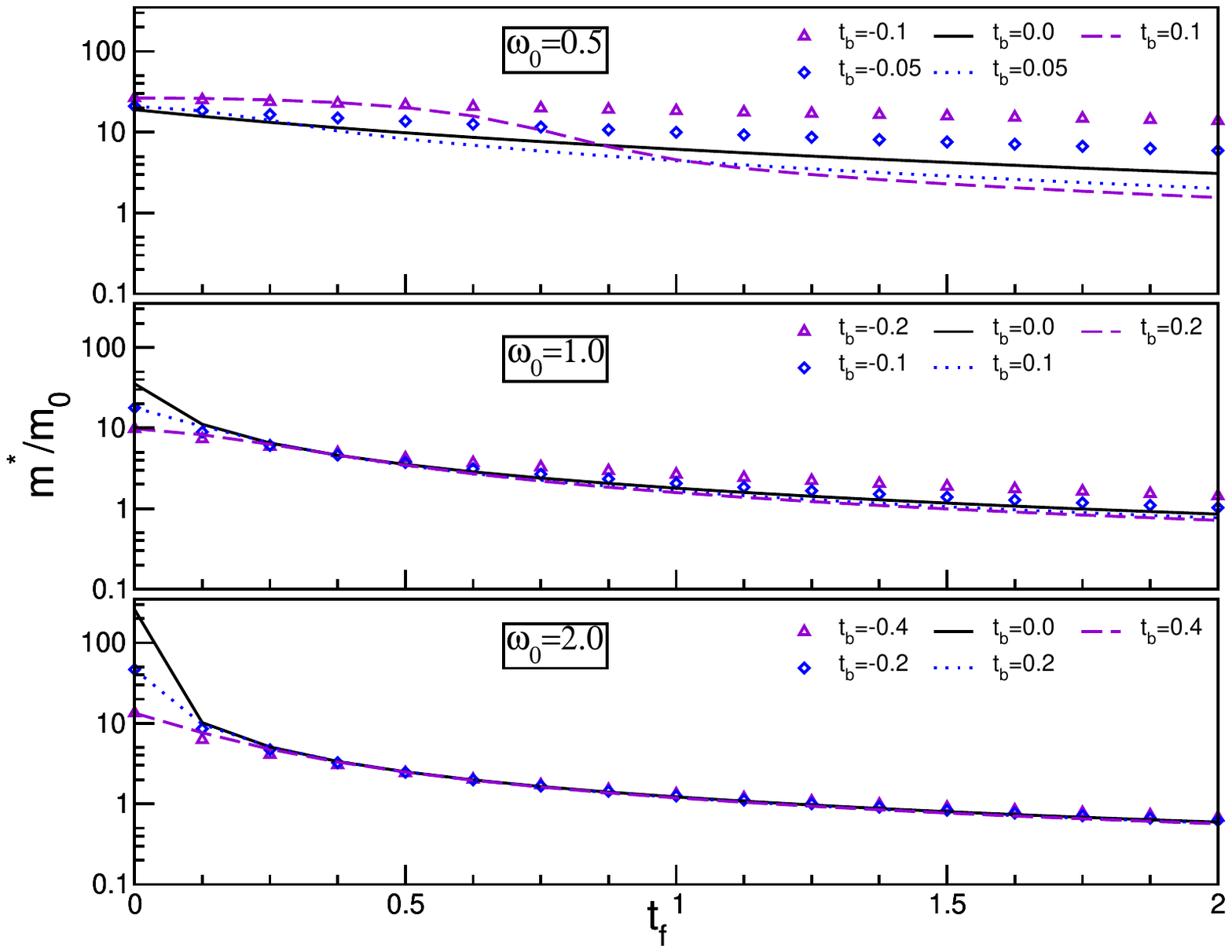} 
\caption{Mass enhancement in the 1D Edwards model with dispersive bosons as a function of $t_f$.}
\label{fig8}
\end{figure}
\begin{figure}[t]
\includegraphics[width=0.95\linewidth]{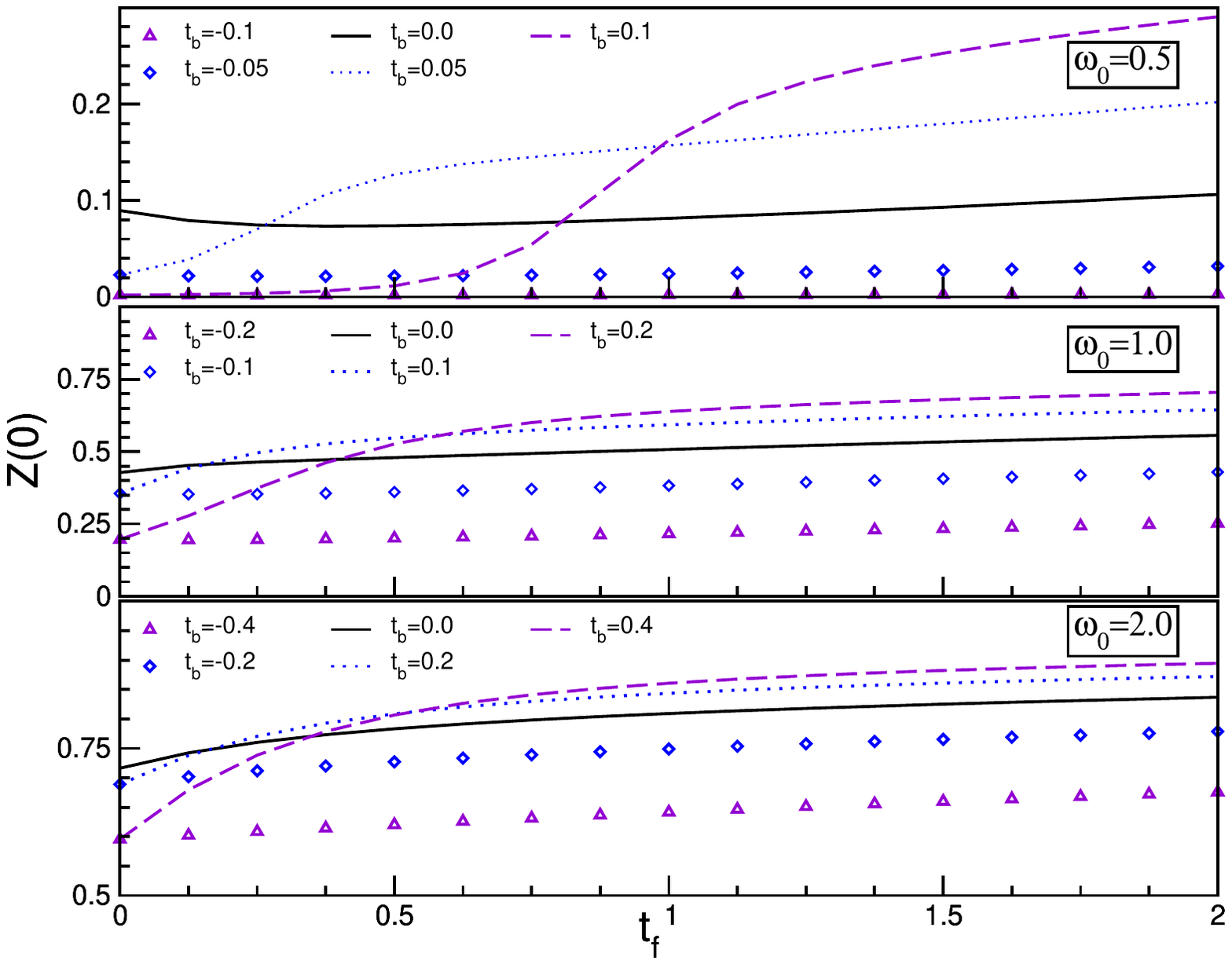} 
\caption{Quasiparticle weight in the 1D Edwards model with dispersive bosons as a function of $t_f$. Note the different ordinates scaling.}
\label{fig9}
\end{figure}
\begin{figure}[h!]
\includegraphics[width=0.95\linewidth]{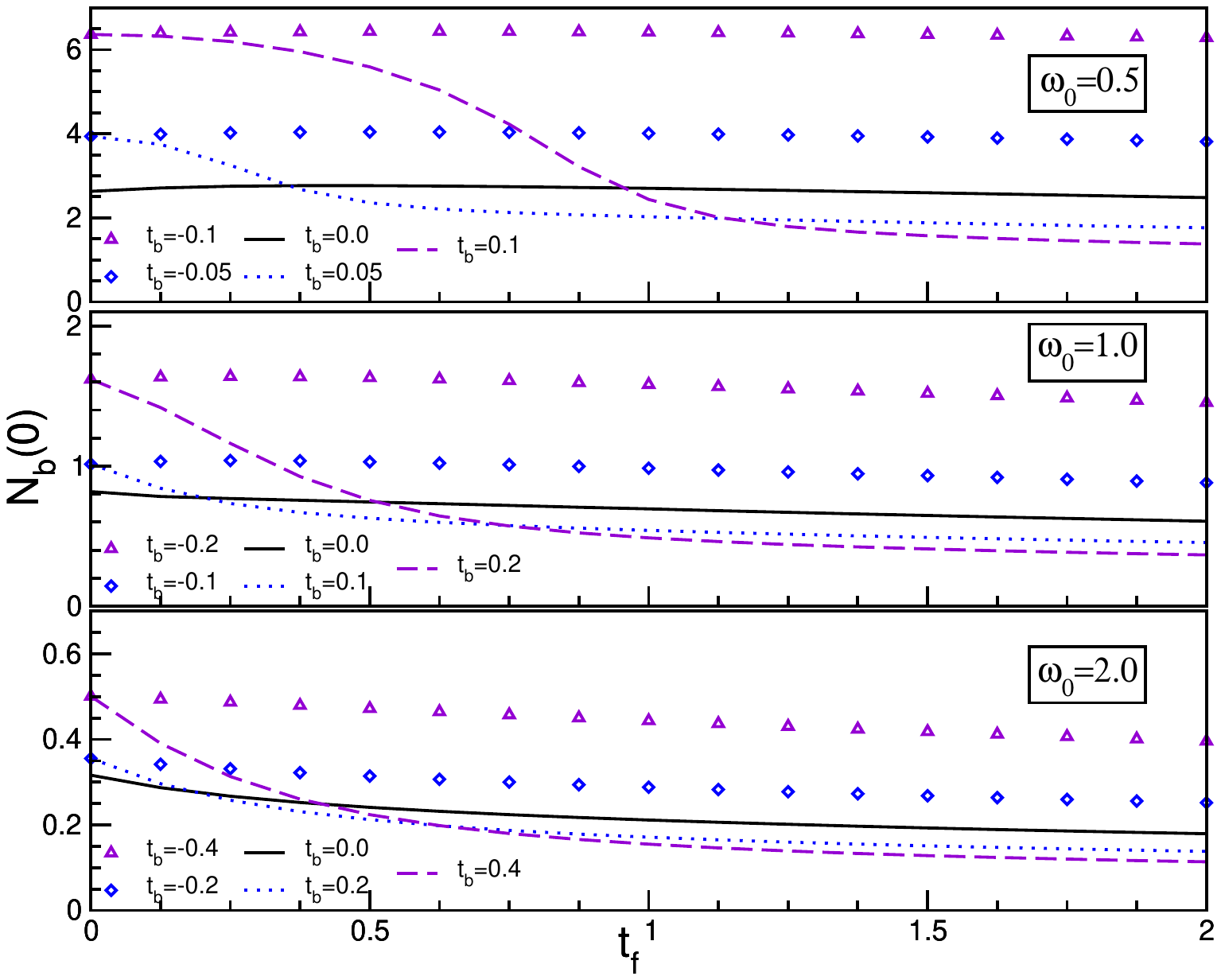} 
\caption{Mean boson number  in the ground state of the 1D Edwards model with dispersive bosons as a function of $t_f$. Take note of the different ordinates scaling.}
\label{fig10}
\end{figure}

This analysis is corroborated by Figs.~\ref{fig8},~\ref{fig9} and~\ref{fig10}, which provide results for the effective mass, the quasiparticle weight, and the mean boson number in the ground state as a function of $t_f$, respectively. Here, the top, middle, and bottom panels characterize the respective behavior in the fluctuation-dominated, crossover, and correlation-dominated regimes. While in the first regime the $t_b$- dependence is significant, it is rather weak in the latter. It is obvious that the strongest mass enhancement takes place at large $\omega_0$, for small values of $t_f$ and $t_b$. Interestingly, the quasiparticle weight exhibits a  stronger dependence on the width of the phonon dispersion than the effective mass, in particular $Z(0)$ is much stronger reduced at small $t_f$ in the fluctuation-dominated region where the bosons act as random scatters. The mean boson number shows a corresponding behavior. As it was to expected, the number of bosons is large in the fluctuation-dominated regime, where the difference between the upward $t_b>0$ and downward $t_b<0$ dispersion becomes especially significant in the quasi-free transport regime of large $t_f$, see top panel in Fig.~\ref{fig10}. Quite different, in the correlation-dominated regime, the number of bosons is small and its dependence on $t_b$ is almost negligible, see bottom panel in Fig.~\ref{fig10}. 

We complete our investigation of the ground-state properties of the modified Edwards model by looking at the particle-boson correlation function
\begin{equation}
  \chi_{ij} = \langle 0| f^\dagger_i f^{}_i  b^\dagger_j b^{}_j  | 0 \rangle \;.
\end{equation}
\begin{figure}[t]
\includegraphics[width=0.95\linewidth]{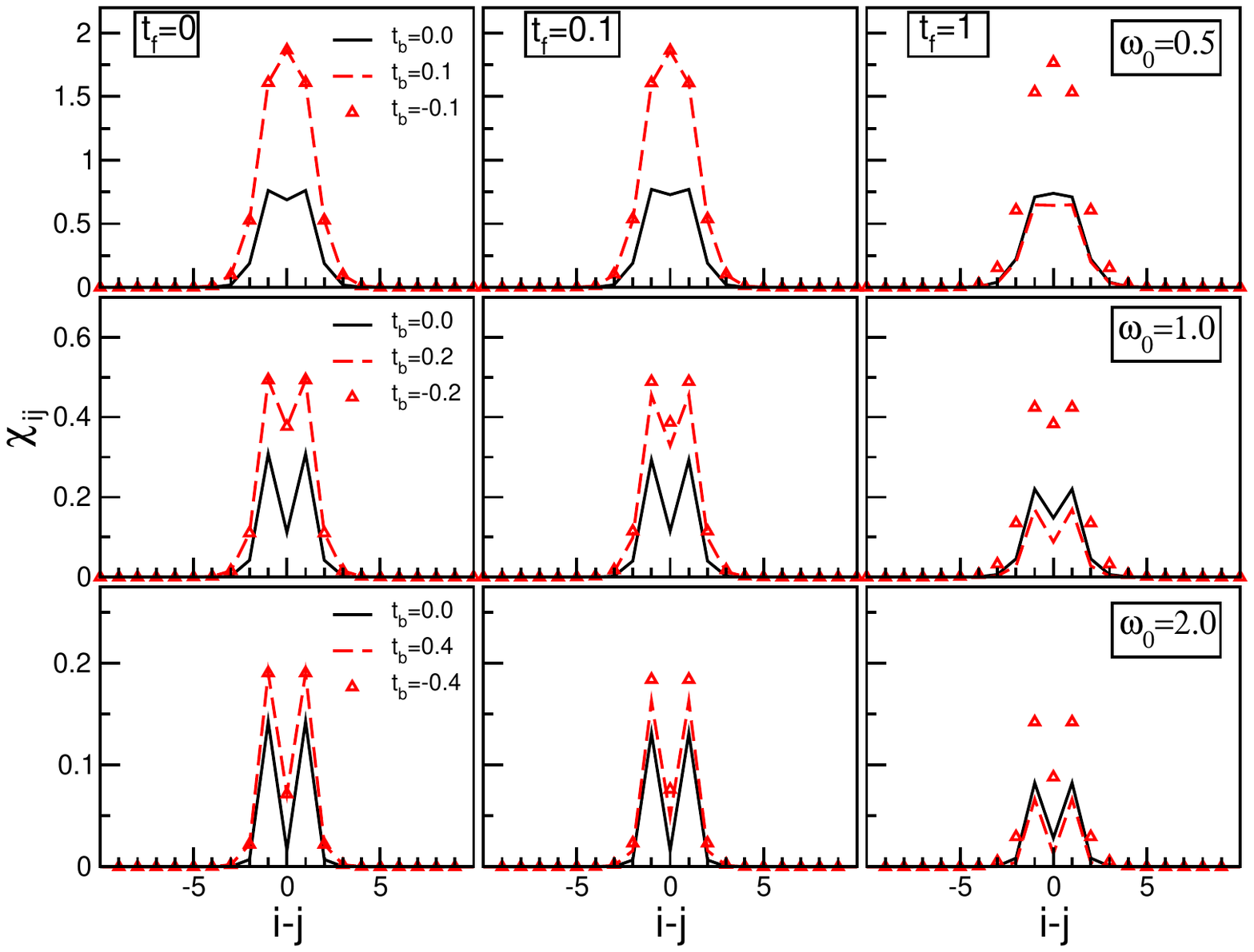} 
\caption{Particle-boson correlation function in 1D (dispersive boson) Edwards model computed at different $t_f$, $t_b$, and $\omega_0$. Note the different ordinate scalings.}
\label{fig11}
\end{figure}
In Fig.~\ref{fig11} we present the spatial correlations between a particle located at site $i$ and a bosonic excitation of the background located at site $j$. Starting from the left bottom corner, 
the left column shows how the strong correlations between the particle and the bosons at the adjacent sites---which is energetically advantageous because the annihilation of an existing  neighboring boson promotes the boson-controlled hopping transport---weaken when the bosons can be excited more easily, i.e., the background increasingly fluctuates, see top left panel.    
Obviously $\chi_{ij}$ is enhanced by a finite $t_b$ in any case, where the  dependence on the sign of  $t_b$ is not very pronounced, however. This tendency increases as $t_f$ increasing; follow the top row from left to right. The upper right panel shows that in the  quasi-free transport regime the bosons of course form a cloud surrounding the particle but are otherwise weakly correlated. In this regime the correlations are much more pronounced for bosons with an upward- bended dispersion  than for those with a downward-bended one, which is understandable because the broader electron band  and the lowered phonon dispersion interfere more easily at zero momentum. If we now increase the boson frequency at $t_f=1$, we can see how the particle-boson correlations develop again; look down the right column. This effect increases when we decrease $t_f$ at fixed $\omega_0=2$, following the last row from right to left. Thereby the dependence of $\chi_{ij}$ on the sign of $t_b$ largely disappears.

\subsection{Spectral properties}
We now turn to investigate the influence of boson dispersion on the 
single-particle spectral function, 
\begin{equation}
\label{aspekt}
A(k,\omega) 
=  \sum_n 
|\langle n|f^\dagger_k 
|\mathrm{vac}\rangle|^2 \,\delta [\omega-\omega_n]  \;,
\end{equation}
which is directly accessible experimentally via (inverse) photoemission. 
In Eq.~\eqref{aspekt}, $|{\rm vac}\rangle$ means the particle vacuum, 
and $|n\rangle$ labels the eigenstate of the one-fermion system
with excitation energy $\omega_n = E_n-E(0)$.    

\begin{figure}[t]
\includegraphics[width=0.95\linewidth]{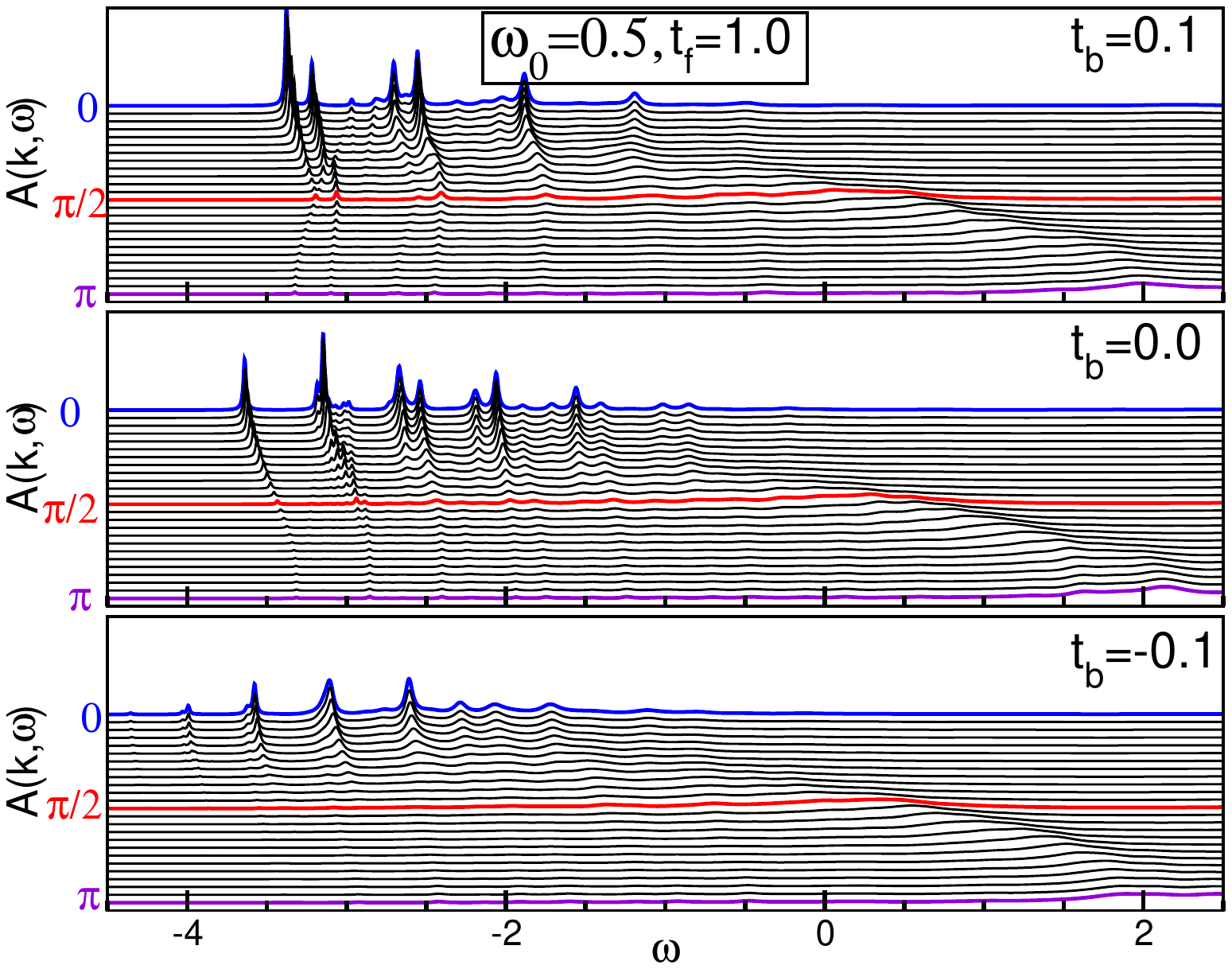} 
\includegraphics[width=0.95\linewidth]{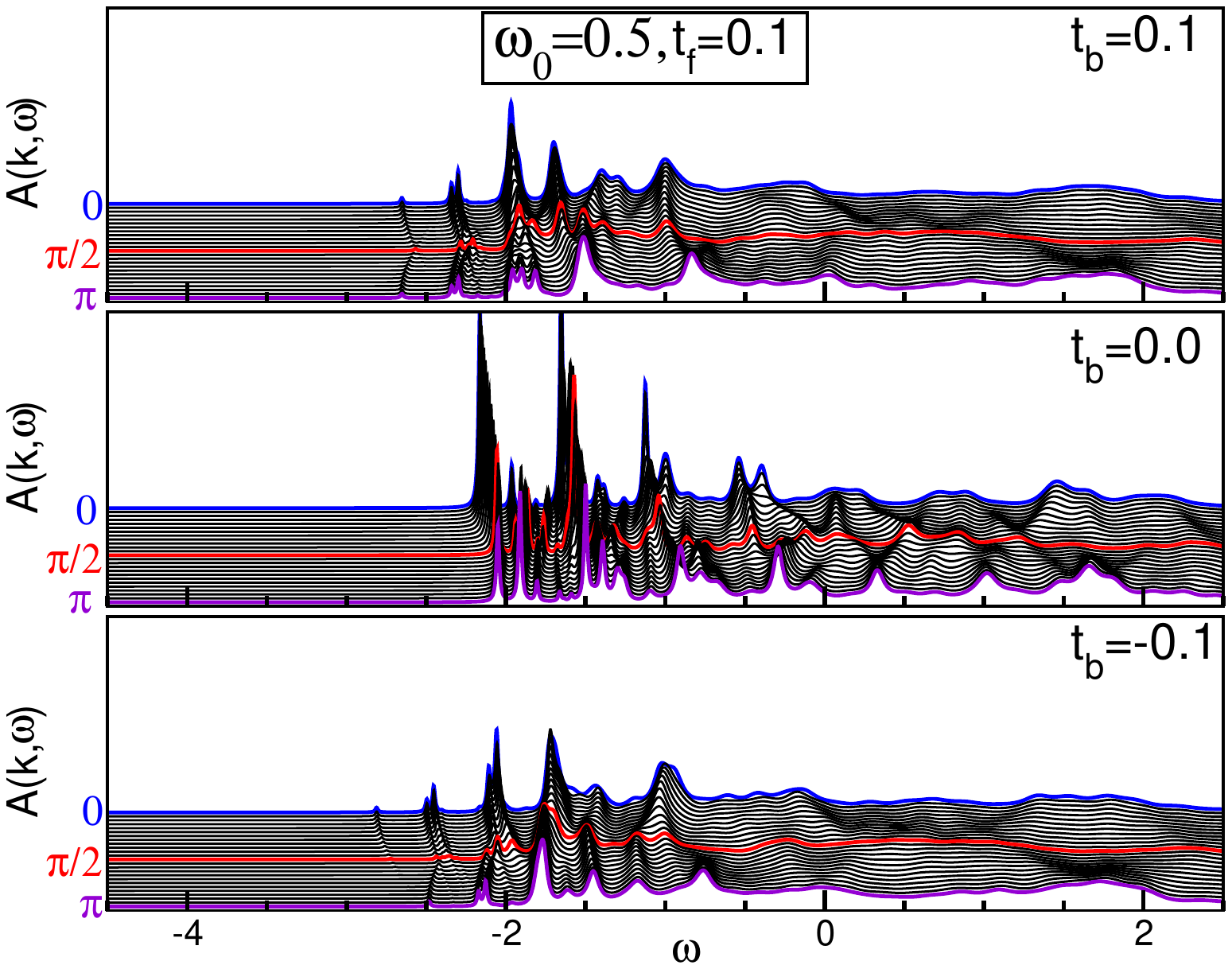} 
\caption{Single-particle spectral function $A(k,\omega)$ in  the 1D Edwards model computed at different boson dispersions parametrized by $\omega_0=0.5$ and $t_b$ as denoted in the legends, where   $t_f=1$ (top panels) and $t_f=0.1$ (bottom panels). Note that the values of $A(k,\omega)$ were multiplied by a factor of 10 to increase the visibility.}
\label{fig12}
\end{figure}

\begin{figure}[t]
\includegraphics[width=0.95\linewidth]{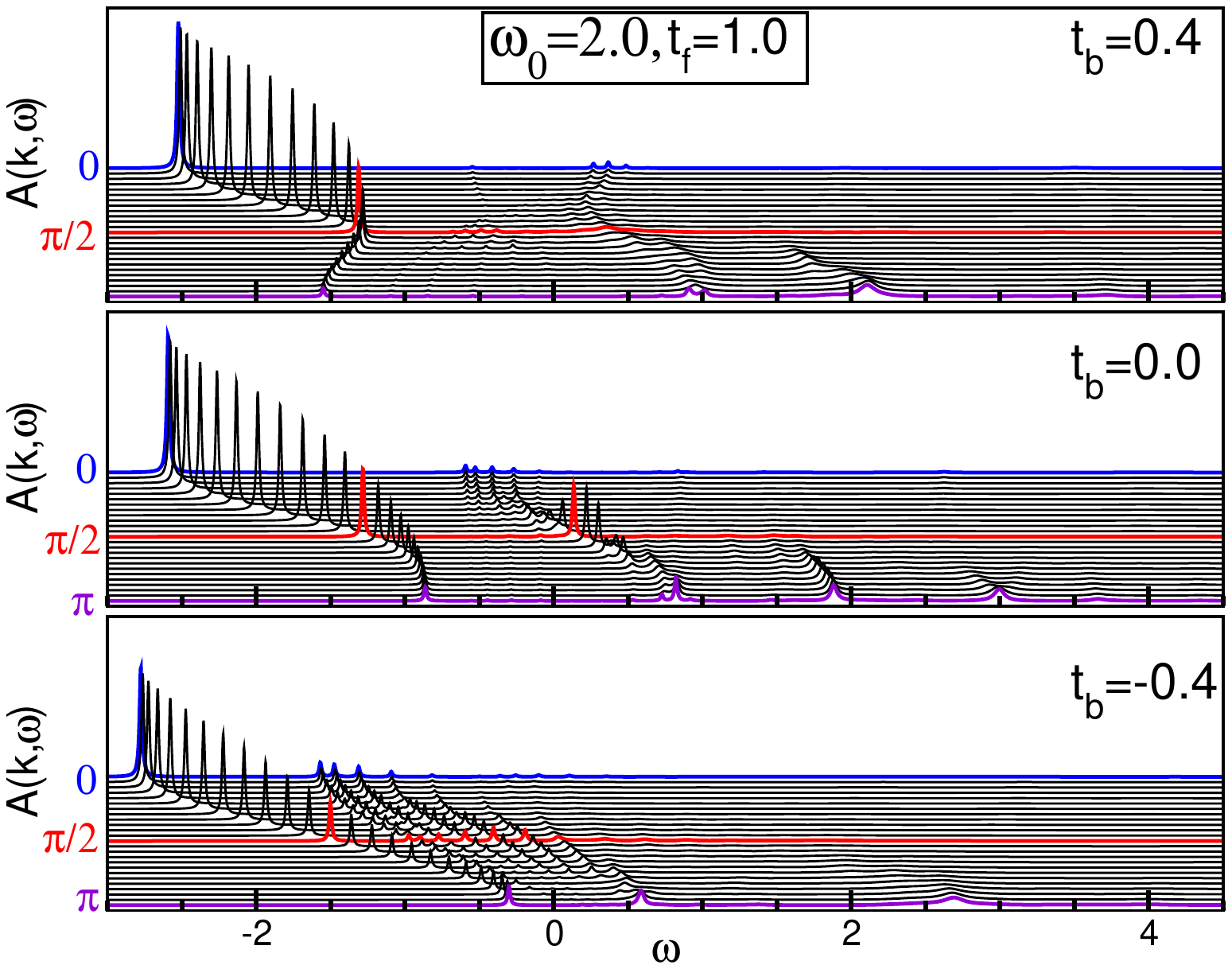} 
\includegraphics[width=0.95\linewidth]{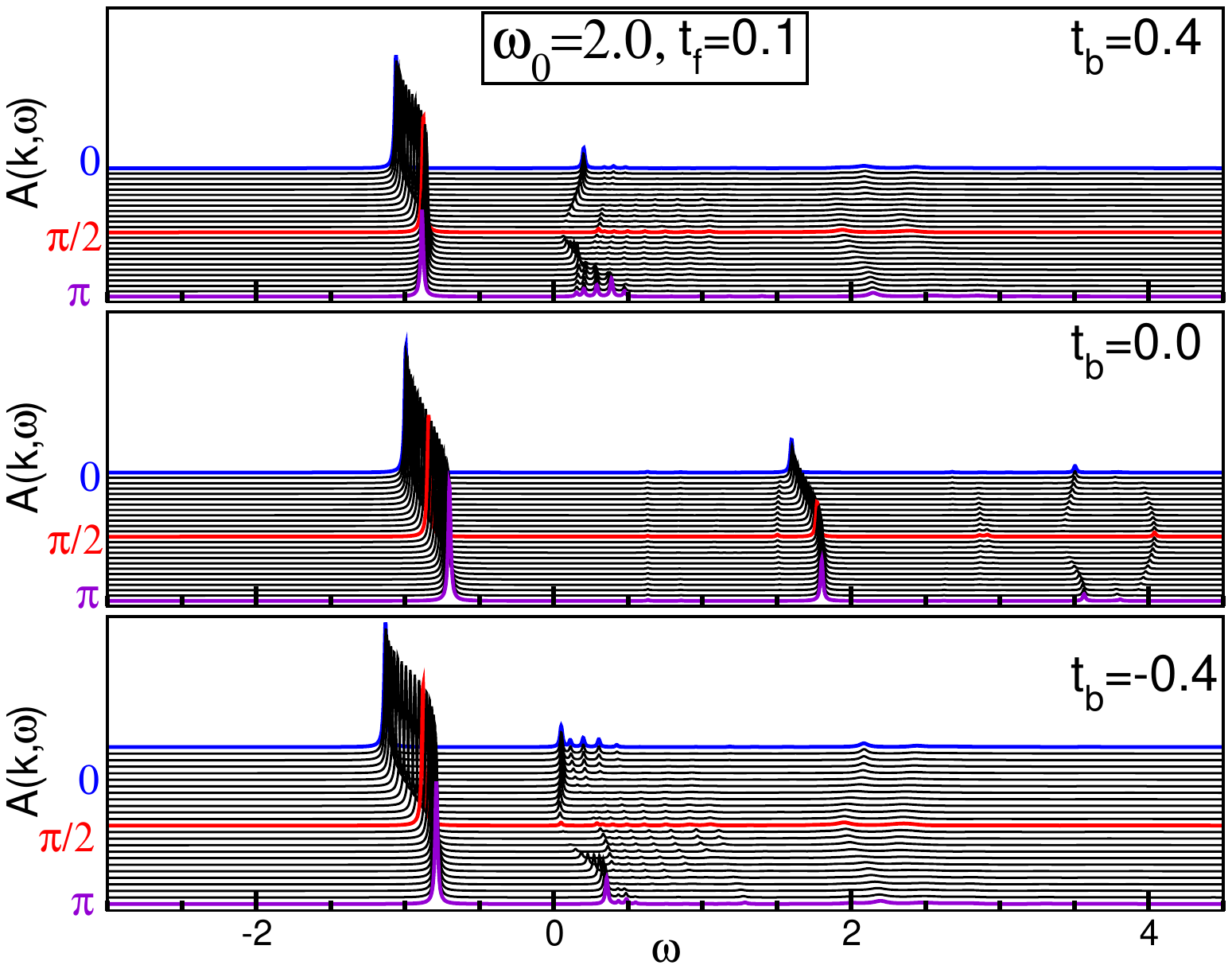} 
\caption{Single-particle spectral function $A(k,\omega)$ in  the 1D Edwards model computed at different boson dispersions parametrized by $\omega_0=2.0$ and $t_b$ as denoted in the legends, where   $t_f=1$ (top panels) and $t_f=0.1$ (bottom panels).}
\label{fig13}
\end{figure}

Figures~\ref{fig12} and~\ref{fig13} present $A(k,\omega)$ in the low- and high-boson-frequency regime, respectively, for representative values of $t_f$ and $t_b$. Here, the middle panels show the results for the pure Edwards model ($t_b=0$) only for comparison. In general, the lowest spectral signature in each panel follows the dispersion of the lowest-energy band from Figs.~\ref{fig4} and~\ref{fig6}, respectively, but we are now able to assign a predominantly electronic or bosonic nature to the excitations at $k$ and $\omega$ via the magnitude of $A(k,\omega)$, more precisely by the wavefunction renormalization factor   
\begin{equation}
Z({k})=\vert\langle \psi_k\vert f_{k}^{\dag} \vert \mathrm{vac}\rangle\vert^{2}\,,
\label{Z}
\end{equation}
where $\psi_k$ denotes the single-particle state with momentum $k$ being lowest in energy.
 It is therefore important to note that, as we move from $k$=$0$ to $k$=$\pi$ in Figs.~\ref{fig12} and~\ref{fig13}, we shift the {\it y}- coordinate associated with a given $k$ value by a fixed amount with respect to that of the previous $k$ value (by $-0.05$ in the case of Fig.~\ref{fig12} and by $-1$ in the case of Fig.~\ref{fig13}), so that the spectral function is well resolved at  each $k$- point.

Again we start the analysis in the fluctuation-dominated regime of relatively small $\omega_0$ and large $t_f$, see Fig.~\ref{fig12} top panels. Obviously there is no well-developed quasiparticle band in this case (see the very weak electronic signature along the ``free'' dispersion $-2t_f \cos k$ between $\omega=-2$ and~2), but a sequence of predominantly bosonic bands separated by $\omega_0$. We furthermore find that the quasiparticle weight is almost negligible away from $k=0$ (remember the amplification factor 10), what means that the particle motion is rather incoherent (overdamped by bosonic fluctuations).   
This is observed for vanishing $t_b$ but retains its validity for (small) finite  $t_b$, whereby the lower ``bosonic'' bands are weaker (stronger) and are lowered more (less) for $t_b<0$ $(t_b>0)$ compared to the case $t_f=0$. This behavior is understandable if one considers the shift of $\omega(0)$ and the curvature of $\omega(q)$ for $t_b {>\atop <}0$ in the course of the crossing of electron and boson bands for a weakly correlated fermion-boson system.

If $t_f$ is reduced ($t_f=1\to0.1$), correlations begin to develop in the coupled fermion-boson system, i.e., the transport becomes more and more boson controlled, even at a comparatively small $\omega_0=0.5$. This is illustrated in the bottom panels of Fig.~\ref{fig12}.  While for $t_f=0$ a relatively strong quasiparticle band is already formed, a finite boson dispersion resulting from  $t_b {>\atop <}0$ still causes rather bosonic bands at very low energies, separated approximately by $\omega_0$. By contrast, a strong mixing of  multi-phonon and electronic excitations takes place at larger energies $\omega$, where the contributions to single-, two-, or multi-bosons excitations can not be distinguished between each other just as for the Holstein model with dispersive phonons\cite{BT21}, but  certain bands excitation are still recognizable. Here, the differences between the cases with positive and negative $t_b$ are less pronounced. 

The situation changes completely if we consider a stiff background, parametrized by large boson energies $\omega_0$, see Fig.~\ref{fig13}. Now transport is fully boson assisted and a coherent quasiparticle band emerges, although on a reduced energy scale which is mainly determined by $t_f$; compare the middle panels, where $t_b=0$, for $t_f=1$ and $t_f=0.1$. Here again the higher excitation bands are positioned at multiples ($m$) of the boson energy $\omega_0$ above the lowest-energy band, whereby their intensity is greatly reduced with increasing $m$. For $t_b=0.4$ ($>0$)  
there is a non-monotonic downward bending of the lowest electron band due to the dispersive bosons, which is absent for $t_b=-0.4$ ($<0$). In the very correlation-dominated regime 
($\omega_0=2,\;t_f=0.1$), interestingly, dispersive structures occur in $A(k,\omega)$ below the single-boson excitation, which are due to two-boson excitations similar to those observed in the (dispersive) Holstein model\cite{BT21}. In the Edwards model case, however, their spectral weight is not only concentrated in a narrow interval around the center of the Brillouin zone but also at its boundary.

\section{Conclusions}
The paradigmatic two-channel Edwards transport model has so far only been studied for dispersionless bosons, just as the  Holstein small polaron model until recently\cite{MB13,BT21}. 
In this work, we were able to show that the inclusion of the dispersive bosons has a profound effect on both the static and dynamic properties of the  
fermion-boson Edwards model in the single-particle sector. 

We demonstrate that, as in the pure Edwards model, the background parametrizing bosons act in two competing ways: They limit transport when they fluctuate strongly and are weakly 
correlated, or they assist transport by boson-supported hopping in the regime of strong correlations where the boson relaxation rate is low. Although the principle regions of dominant quasi-free, diffusive and boson-assisted transport of the Edwards model  continue to exist naturally, their boundaries shift because the bosons at different sites are no longer independent of each other. 
While a finite boson dispersion  generally increases the effective mass (decreases the Drude weight) of the particle in the fluctuation-dominated regime, the effective mass is reduced, i.e., the particle becomes more mobile, in the correlation-dominated regime, where the varying effect of concavely and convexly curved boson dispersions is rather small. 

The influence of a finite upward and downward curved boson dispersion is best reflected in the single-particle spectral function, which allows statements about the position, shape and strength of the quasiparticle band, as well as the nature of the excitations. In the fluctuation-dominated regime, around $k=0$ and $k=\pi$, the (over)damped character of the particle motion becomes visible.  The quasiparticle band is most pronounced in the quasi-free and correlation-dominated regimes, where the bandwidth is strongly reduced in the latter case. Perhaps unexpected is the observation of dispersive spectral weight between the lowest-energy band and the single-boson excitation  which can be attributed to multi-boson excitations\cite{BT21}. 

This suggests that any quantitative modeling of the transport properties of materials exhibiting lattice or spin polaron physics using the Edwards fermion-boson model must take into account the dispersion of the bosons parametrizing the background medium. Especially the  multi-boson excitations found in the electronic spectral function will also noticeably influence the nonequilibrium and finite-temperature properties  and should be detectable experimentally already at small boson dispersion\cite{BT21}.

From a theoretical point of view, our results will certainly stimulate further investigations, for example with respect to the influence of the boson dispersion on the Tomonaga-Luttinger-liquid charge-density-wave quantum phase transition, which has been proved to exist in the pure 1D Edwards model at half band filling\cite{EHF09}.

\acknowledgements  
H.F. acknowledges the hospitality of the Los Alamos National Laboratory where part of this work was done.

%

\end{document}